\documentclass[aps,pre,showpacs,showkeys,twocolumn,groupedaddress]{revtex4}

\usepackage{graphicx}
\usepackage{dcolumn}
\usepackage{bm}

\begin{document}

\title{Modeling gene's length distribution in genomes} 

\author{Stanis{\l}aw Cebrat$^1$, Miros{\l}aw R. Dudek$^2$ and Pawe{\l} Mackiewicz$^1$}
\email{cebrat@microb.uni.wroc.pl, mdudek@proton.if.uz.zgora.pl, pamac@microb.uni.wroc.pl}
\affiliation{$^1$ Division of Genomics, University of Wroc{\l}aw, 
Przybyszewskiego 63/67,\\
$^2$ Institute of Physics, Zielona G{\'o}ra University,
 65-069 Zielona G{\'o}ra, Poland
}

\begin{abstract}
We show, that the specific distribution of gene's length, which is observed in natural genomes,  might be a result of a growth process, 
in which a single length scale $L(t)$ develops that grows with time as
$t^{1/3}$. This length scale could be associated with the length of the longest gene in an evolving genome. The growth kinetics of the genes resembles the one observed in physical systems with conserved ordered parameter. We  show, that in genome this conservation is guaranteed by compositional compensation along DNA strands of the purine-like trends introduced by genes. The presented mathematical model is the modified Bak-Sneppen model of critical self-organization applied to the one-dimensional system of $N$ spins. The spins take discrete values, which represent gene's length.
\end{abstract}

\pacs{82.39.Pj, 89.75.Da, 87.15.Aa, 89.75.Fb}
\keywords{DNA, domain growth, Bak-Sneppen model, computer simulation genome, gene length}

\maketitle

\section{Introduction}
\begin{figure}
\includegraphics[scale=0.3]{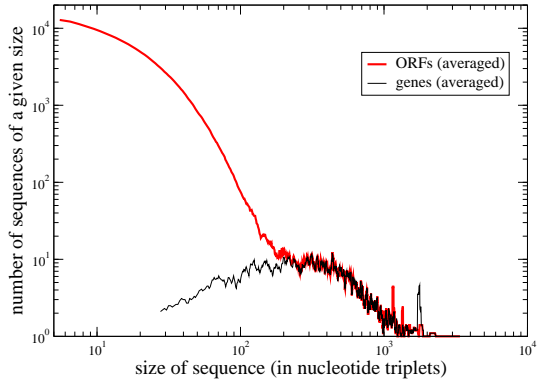}
\caption{Distribution of gene/ORF sizes in the \textit{Saccharomyces cerevisiae} genome. There are 5850 genes and 281361 ORFs. 
The continuous lines, which are plotted for the better data presentation, represent the averaged number of ORFs/genes in classes of the length of $10$ triplets.}
\label{fig_ORFs}
\end{figure}

Can we model main features of the biological processes leading to the gene's length distribution in natural genome? The problem still is open. We do not even know what the processes are.
In Fig.~\ref{fig_ORFs}, we have shown an  example of the gene's length distribution in \textit{Saccharomyces cerevisiae} genome together with ORF's (\textit{Open Reading Frame})  length distribution. Generally, any sequence starting with codon \textit{ATG} and ending with one of three
stop codons: \textit{TAA}, \textit{TGA} or \textit{TAG} is called ORF. Here, the symbols \textit{A}, \textit{T}, \textit{G}, \textit{C} denote DNA nucleotides.
In practice, the ORFs consisting of more than $k=100$ nucleotide triplets (one nucleotide triplet codes for one amino-acid) 
are considered to be coding, whereas short ORFs are considered to be random sequences.
The gene's and ORF's size distribution share common properties in all natural genomes.  It is worth emphasizing that the DNA replication and
transcription processes have different organization in prokaryotic organisms 
and eukaryotic organisms. In prokaryotes, there exists only one region of the origin of replication (ORI), the role of differently replicating DNA strands (called leading and lagging) is established and the transcription is closely related to the replication process, whereas in eukaryotes, there are many sites of the origin of replication, the role of DNA strands may be changed in successive replication cycles and the transcription occurs at different times than the DNA replication.
Moreover, there exist the different distributions of coding sequences on chromosomes in both types of organisms. The majority of genes are preferably located on the leading strand in prokaryotes whereas in eukaryotes we can find regions of chromosome that are rich and poor in genes usually related to the isochoric organization of chromosomes. 
Genes might have different origination, different arrangement on DNA sequences and they also might be coupled in functional clusters  (e.g., see review paper by Wolfe and Li \cite{WolfeLi}).
 It is accepted that the observed long-tail structure in length distribution of ORFs, as in Fig.~\ref{fig_ORFs}, results  from the direct selective pressure (e.g., \cite{WentianLi},\cite{Cebrat1},\cite{proteomesize}).  

Li et al. \cite{WentianLi2} have studied the statistical properties of  
\textit{Saccharomyces cerevisiae} genome and they have observed   the compositional non-randomness in this genome at large length scales including very long and very short genes.  In particular, the authors concluded that long genes (ORFs) cannot have a random origin.

There was the long-time discussion, between physicists, mathematicians and geneticists, on long-range correlation of nucleotides along DNA strands, which was started in 1992 by  Li \cite{Li_correlation}, Peng et al. \cite{Stanley_correlation}, and Voss \cite{Voss_correlation}. At the beginning, it was stated that the long-range correlation is observed only in non-coding sequences, whereas coding sequences resemble random sequences, with one exception:  the evidence of triplet structure of genetic code.  Later on, it was shown that also coding sequences can be  correlated   (\cite{Arneodo_correlation1}, \cite{bacteria_correlation}). Recently, Vaillant et al. \cite{Arneodo_correlation2}  have suggested that the existence of the long-range correlation up to the distances 20 -- 30 kbp can be connected with the nucleosomal structure and dynamics of the chromatin fiber . 

There is a very spectacular property of genes, that if all coding sequences are taken together, they appear as compositionally random sequences in the length of scale of the whole genome (white noise power spectrum in Fourier analysis).
In the following two subsections, we will show, that the reason for this behavior is, that the coding sequences try to compensate purine-like bias introduced by nucleotides they are composed of. The compensation takes place both with the help of other coding sequences and also with non-coding sequences. Some properties of this finding were already published in 1996 \cite{Proceedings} and now we come back to it. We will also use the Jensen-Shannon divergence \cite{Bernaola1} to show this compensation property as entropic uniformity of DNA coding sequences.

In the paper by Dembska et al. \cite{Dembska}, in which  the possible effect of food-chain correlation on nucleotide fraction of competing species was considered, the authors  suggested that  genes in a genome might experience similar self-organization mechanisms as the species in the Bak-Sneppen model  of evolution \cite{BS1}. 
If this this is correct, the location of genes in genome and gene's size  represent a process, which is very far from the equilibrium with events 
corresponding to the avalanches phenomenon in the Bak-Sneppen model \cite{BS1}. We have followed this idea and introduced a spin model representing genes on two DNA strands. In this model the distribution of spin size shares many features common with the distribution of ORFs and genes in natural genomes. In particular, two distinct parts in the size distribution, for small ORFs and for long ORFs (genes) may be distinguished, as in Fig.~\ref{fig_ORFs}. We have shown, that the tail-like part of this distribution, representing genes, develops in time as $t^{1/3}$. 

\subsection{Symmetry triplet and anti-triplet}
\begin{figure}[t]
\includegraphics[scale=0.3]{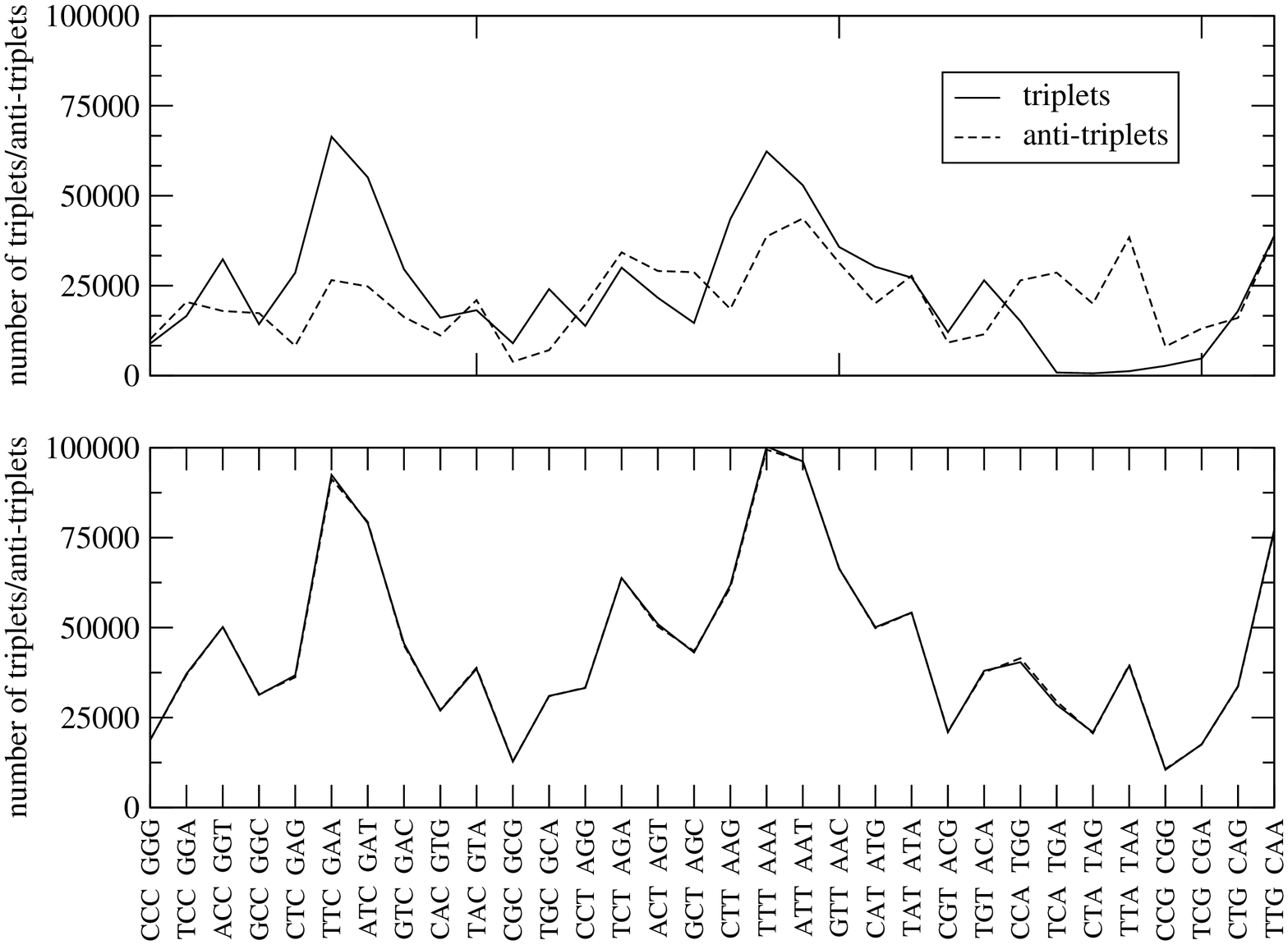}
\caption{Joint histogram of the usage of  triplet and anti-triplet in all 16 chromosomes of  \textit{Saccharomyces cerevisiae} genome: the upper part of the figure is solely for ORFs ($ \ge 150$ triplets) on the strand W of these chromosomes, whereas the bottom part has been made both for ORFs ($ \ge 150$ triplets) on the strand W, as above, and the  antisense of  ORFs ($ \ge 150$ triplets) from strand C, i.e. their triplets are read on W in direction of W, e.g., for triplet ATG on C  strand its antisense is read as CAT on W strand.}
\label{fig_YeastTriplets}
\end{figure}
\begin{figure}[t]
\includegraphics[scale=0.3]{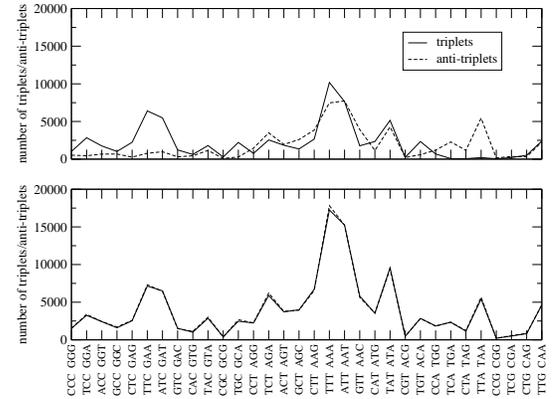}
\caption{Histogram of using  triplet and anti-triplet of ORFs  ($ \ge 150$ triplets) in \textit{Borrellia burgdorferi} genome. The meaning of the upper part of the figure 
and the bottom part is the same as in Fig.~\ref{fig_YeastTriplets}.}
\label{fig_BorrelliaTriplets}
\end{figure}

The observation of compositional uniformity of coding sequences in DNA strands is closely connected with the expectation that in ideal genome numbers of nucleotides should satisfy the balance condition, $A=T$ and $G=C$, in each DNA strand. A short review of the experimental data and comments on them can be found in the paper by Lobry and Sueoka \cite{LobrySueoka}.
How strong the trend is can be seen in Fig.~\ref{fig_YeastTriplets}  (some results have been published in \cite{Proceedings}), where we have plotted the distribution of nucleotide triplets used by long ORF,  longer than 150 triplets, in the \textit{Saccharomyces cerevisiae} genome. We might deduce from Fig.~\ref{fig_ORFs}  that almost all these ORFs are  genes.
As in \cite{Proceedings}, we have divided all 64 triplets into two groups: the group rich in purines and the other rich in pyrimidines. We called them triplets and anti-triplets, respectively. The bottom part of Fig.~\ref{fig_YeastTriplets} is the most interesting one, because it shows the strong triplet and anti-triplet symmetry in the nucleotide content of genes if the coding information is read in one DNA strand only (Watson or Crick).  It is evident that the genes of Watson strand try to compensate genes of Crick strand. This finding is non-trivial because the analyzed genes do not overlap! This means, that the distinct DNA regions do compensate with each other. Thus,  genes occupy the positions in which the trends introduced by them  are compensated on the same strand by the sequences complementary to the genes of the opposite strand. 
In the \textit{Saccharomyces cerevisiae} genome, the same triplet and anti-triplet symmetry  holds on many legths of scales in each of 16 chromosomes \cite{Proceedings}.   This is not the case in bacterial genomes, like in Fig.~\ref{fig_BorrelliaTriplets}, where the symmetry can be observed only on the length scale of the entire genome.

\begin{figure}
\includegraphics[scale=0.3]{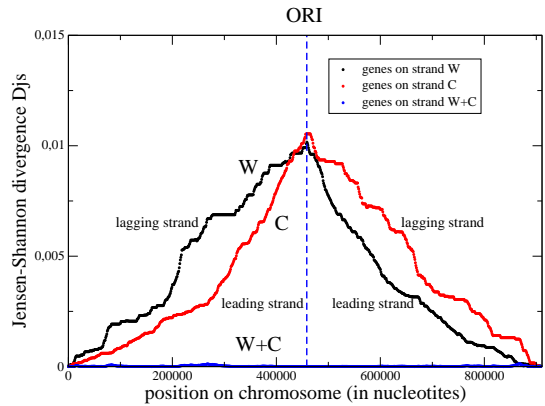}
\caption{Dependence of the value $D_{js}$ on position on the DNA sequence  of the \textit{Borrellia burgdorferi} chromosome, in which the DNA sequence is partitioned into two segments, left- and right-side  segment. The larger value of $D_{js}$  the larger the compositional differences in the left and right segment. Here, the largest value of  $D_{js}$ is at the ORI location (458476 bp).}
\label{fig_BorrelliaEntropy}
\end{figure}

\begin{figure}
\includegraphics[scale=0.3]{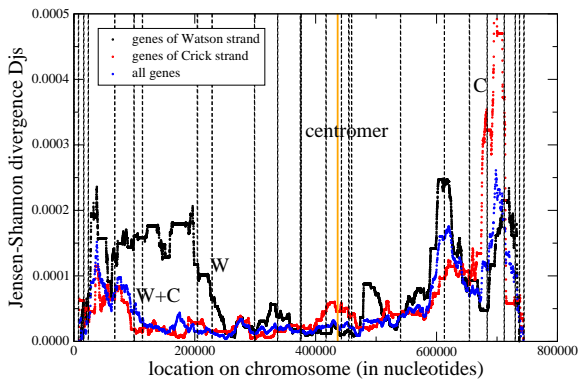}
\caption{The same as in Fig.~\ref{fig_BorrelliaEntropy} but for chromosome X of the \textit{Saccharomyses cerevisiae} genome. The vertical lines represent ARS location (\textit{Autonomously Replicating Sequence}).}
\label{fig_YeastEntropy}
\end{figure}

\subsection{Jensen-Shannon divergence}

The specific compositional compensation of genes discussed in the above subsection can be observed also with the help of entropy.
We have calculated the Jensen-Shannon divergence \cite{Bernaola1}, $D_{js}$, along DNA strands of the genomes under consideration. The two particular examples, one for prokaryote and one for eukaryote, are shown in  Figs.~\ref{fig_BorrelliaEntropy} and \ref{fig_YeastEntropy}, respectively.

The Jensen-Shannon divergence has been used 
in the segmentation algorithm described first in Bernaola-Galv{\'a}n et al. \cite{Bernaola1} to detect compositionally different regions of DNA.
In this method a DNA sequence, which is $L$ nucleotides long is partitioned into two segments at some nucleotide position $i$ ($1 \le i \le L-1$) and next,  
the Shannon entropy is calculated both for these two segments and  the entire DNA sequence. The Jensen-Shannon difference at the position $i$ is defined as follows
 
\begin{equation}
D_{js}(i)=H(p_1\,F_1+p_2\,F_2)-p_1 H(F_1)-p_2 H(F_2), 
\end{equation}

\noindent
where $H(F_1)$ and $H(F_2)$ represent the Shannon entropy for nucleotide fraction $F_1=\{F_A^{(1)},F_T^{(1)},F_G^{(1)},F_C^{(1)}\}$ of the segment (1) and nucleotide fraction 
$F_2=\{F_A^{(2)},F_T^{(2)},F_G^{(2)},F_C^{(2)} \}$ of the segment (2), and $p_1=i/L$, $p_2=(L-i)/L$ are the weight factors for the segment (1) and (2), respectively. The entropy is defined as follows

\begin{equation}
H(F)=-\sum_{k=A,T,G,C}F_k\,\log (F_k). 
\end{equation}

We have used the above method of the entropy difference with a small modification, that the nucleotide fraction $F_k$ was restricted to genes only and not to the entire DNA strands as in \cite{Bernaola1}. Thus,  in Figs.~\ref{fig_BorrelliaEntropy}--\ref{fig_YeastEntropy} only the heterogeneity of coding sequences was hunted for.  We have obtained similar results for other bacterial genomes, like \textit{Escherichia coli} genome, \textit{Mycoplasma genit.} etc. However, we have not studied eukaryotic genomes other than the \textit{Saccharomyces cerevisiae}. 
It is interesting to notice in these figures that the compositional heterogeneity introduced by leading and lagging DNA strands (enhancement at the ORI region) practically dissapears in the case when both genes from Watson strand and Crick strand are taken into account. 
In Fig.~\ref{fig_BorrelliaEntropy}, the curve W+C practically lies on the abscissa
on the length scale used in  the figure.
 A similar trend can be observed in Fig.~\ref{fig_YeastEntropy}, in which genes from Watson 
and Crick DNA strands try to compensate any heterogeneity introduced by them. The observed asymmetry is evident only in  the telomeric regions of chromosome. 

The presented examples confirm the discussed above triplet and anti-triplet symmetry of the coding sequences in the entropy language.

\section{Genes as  a spin model}
In the Introduction, we have presented two observations of compensation of the gene's compositional asymmetry along each DNA strand. This property is analogous to the condition, $A \sim T$ and $G \sim C$, in each DNA strand of a natural genome. 

If we take into account the specific purine-like bias in gene's nucleotide composition, we could agree that in the zeroth order approximation the genes could be treated as rigid rods.  In addition, if we make the distinction between genes of Watson and Crick strands assigning a different direction to each rod, we obtain a model of spins represented by arrows directed
up and down. The spin size is represented by the length of gene counted in nucleotide triplets. In this simplified model, the gene's compensation takes place if its neighboring spins have the same absolute value but the opposite sign (direction) or a gene is compensated by a few smaller genes - spins with the opposite sign. Thus, we forget about all the details concerning nucleotides. Only the gene's length is discussed.

In our model,  the magnitude of compensation of each gene is considered to be its fitness parameter in the genome. If $S_i$ denotes spin $i$, which may take both positive and negative value, the fitness parameter of the gene $i$, represented by this spin, is defined as follows:

\begin{equation}
B_i=1-\vert {\sum_{k=i-n}^{i+n} S_{k}} \vert / \sum_{k=i-n}^{i+n} \vert S_{k}. \vert
\label{fitness}
\end{equation}

\noindent
It takes values from the unit interval $[0,1]$ and $n$ represents the number of left-hand and right-hand neighbors of spin (gene) $i$. Notice, that if all spins in the neighborhood of $S_i$ had the same sign as $S_i$ then $B_i=0$. We have $B_i=1$ only if all $2 n+1$ spines are compensated. Notice also that in the formula in Eq.(\ref{fitness}) there is no direct dependence on the absolute value of the spins but the compensation of incompatibility is only taken into account.

In natural genome, genes code some function of an organism. They experience the mutation and selection pressure.  
The genes may be eliminated, adapted to new nucleotide arrangement or duplicated. They can even jump between DNA strands (\cite{jumping1}--\cite{jumping2}) to survive selection. Thus, the distribution of gene's length in the natural genome is the result of a complex process but not the result of the static property of the coding demands only. 
As we have mentioned in the Introduction, we have chosen 
 the Bak-Sneppen \cite{BS1} dynamics as a candidate for modeling such a  process.
The original Bak-Sneppen model describes the co-evolution of  $N$ species, where the species are represented by sites ($i=1, 2, \ldots, N$) of linear chain. The sites are assigned a value $B_i$  from the unit interval $[0,1]$, which measures the surviving fitness of the species $i$. The dynamics of the model is based on a very simple rule that in every discrete time step $t$, the species $i$ with minimum fitness $B_i$ is looked for and  next the species $i$ is replaced by a new one together with the species from the nearest-neighborhood of $i$  - all these sites are assigned new fitness.

\begin{figure}
\includegraphics[scale=0.3]{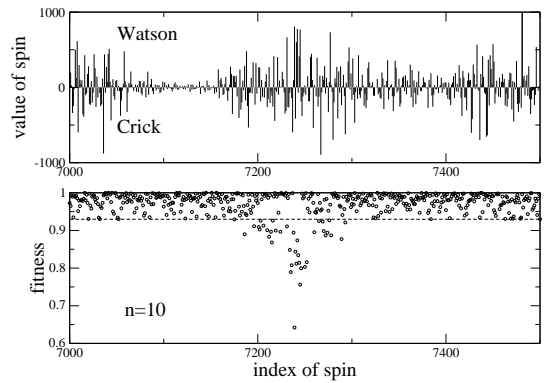}
\caption{Snapshot at time $t=50 \times 10^6$ MC of the avalanche region in  spin configuration (the upper part) and  the same region for spin fitness in the case of adapting neighborhood with  $n_0 \leq n=10$. At this time moment $<n_0 > =2.41$, where the brackets denote arithmetic mean for $N=10 000$ spins. The horizontal dashed line represents global fitness, which is equal $B_c=0.9298$ at  $t=50 \times 10^6$ MC. The value $p/T=0.004$. Initially, all $N$ spins were assigned random $ \pm 1$ values.}
\label{snapshot_spin}
\end{figure}

We have adapted some features of the Bak-Sneppen dynamics to 
our spin model. The fitness condition has been defined in Eq.(\ref{fitness}). We should bear in mind, that in our model the spin size  represents the gene's length and by this it has an integer value. Therefore, we might expect a degeneracy in the number of the spins with the same value of  the smallest fitness $B_i$. If it happens,  we remove at the same time step all these genes together with their surrounding and replace them with the  new ones. In spin language,  this means that the new value,  $S_{new} $, is substituted for the old one. 
However, contrary to the original Bak-Sneppen model, we  change the spin value in a continuous-like way, i.e.,

\begin{equation}
S_{new}=S_{old}+\Delta S, 
\label{deltaS}
\end{equation}

\noindent
where $\Delta S$ is chosen randomly from the given interval $[-D,D]$ and $D$ takes a small value in our simulations. Notice, that this
equation describes the growth kinetics if $\Delta S$ takes small values. 
For large values of $\vert \Delta S \vert $ the spins are assigned new random values as in the limiting case of  the Bak-Sneppen model, i.e., $S_{new}=\Delta S$. In Eq.(\ref{deltaS}), we determine the value of $\Delta S$  with the help of the Metropolis version of the Monte Carlo algorithm as follows:

\begin{itemize}
\item[(i)]  determine a new value of spin $S=S_{old}+D - 2 \gamma D$, where $\gamma$ is a standard random deviate from unit interval,
\item[(ii)] if$(\vert S \vert - \vert S_{old} \vert < 0)$ then accept the new value of spin, $S_{new}=S$,\\
			else take the new random deviate $\gamma$  and
			if $(\gamma < \exp(-\frac{p}{T}(\vert S \vert - \vert S_{old}\vert  ))$
			then accept $S_{new}=S$, otherwise $S_{new}=S_{old}$.
\end{itemize}

\noindent
Here,  $p$ plays the role of tension coefficient and  the temperature $T$ determines a noise level. In the model, the tension could be interpreted as the result of compromise between selectional and mutational pressures leading to  the specific purine-like compositional bias of gene, which should be compensated somewhere along the DNA strand.
On the other hand, the longer gene the larger the probability of its mutation. Therefore, if the ratio $p/T$ is large  the new length $L= \vert S_{new} \vert$  will be decreased rather than increased. 
Hence, we are left with only three parameters, which control the spin value (gene's length): the fitness, $B$, the number of the nearest neighbors, $n$, in the meaning of the Bak-Sneppen model \cite{BS1}, and the reduced tension coefficient $p/T$. 

The whole algorithm, used in our simulations, is the following:

\begin{enumerate}
\item Initially, all spins $S_i$ ($i=1, \ldots, N$) take random integer values from the given interval $[-D,D]$, e.g., $D=1$. There are  applied cyclic boundary conditions, i.e., $S_{N+1}=S_1$.
\item Look for the spins $S_i$ with the smallest fitness $B_i$, where 
$B_i$ is calculated for the self-adapting neighborhood.
 The self-adapting nearest neighborhood of each spin $S_i$  is determined by  such number of $2 n_0$ spins ( $n_0 \le n$), that  satisfy the best compensation of $S_i$ within the range of $2 n$ neighbors. If all $2 n $ spins of the nearest neighborhood of $S_i$ are pointing out  the same direction, then  the value of $n_0$ is set to $n_0=n$. 
\item Change spin values from $S_{old}$ to $S_{new} $ of all spins  which have been found with the smallest fitness as well as the $2 n_0$ spins  from the nearest neighborhood of them,  
according to the kinetic equation in Eq.(\ref{deltaS}).
\end{enumerate}

\noindent
Notice that even if initially all spins take  random small values, then once they have random sign,  they produce large linear tension, proportional to  $B=\sum_i^N B_i$. In the model, this tension can be reduced  with the help of  Eq.(\ref{deltaS}) applied only to these spins which have got the smallest fitness. This simple process producess a global threshold fitness $B_c$ which separates spins with the fitness above $B_c$ and spins with the fitness below $B_c$. The larger value of $B_c$ the stronger the competition between spins. One can observe evolution of the fitness $B_i$ ($i=1, \ldots, N$) of individual spins, which is analogous to the one in the Bak-Sneppen model of species evolution \cite{BS1}. This could also be deduced from Fig.~\ref{snapshot_spin}, where the snapshot of an active avalanche has been shown for the configuration of $500$ spins located in this region  
together with their fitness values $B_i$. In this case the simulated system consists of $N=10 000$ spins which initially took  random values  $\pm 1$. Interestingly, the optimum average number of left (right) neighbors,  $<n_0> \approx 2.4$ is much smaller than $n$ in the simulations performed for $n=10$. 

\begin{figure}
\includegraphics[scale=0.3]{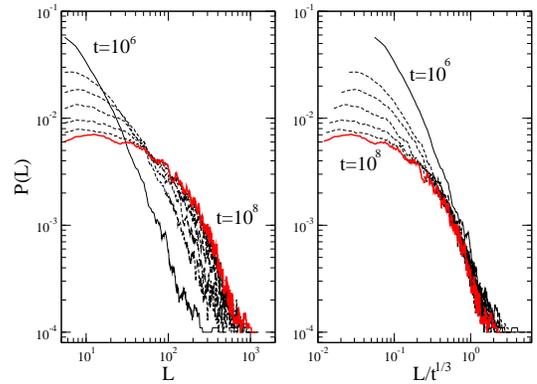}
\caption{Distribution of absolute values of spin at different discrete time steps $t=1 \times 10^6, 1 \times 10^7,  2 \times 10^7, 4 \times 10^7, 6 \times 10^7,  8 \times 10^7, 1 \times 10^8$MC,  when number of spins $N=10000$. The right-hand part of the figure is the same as in the left part but spin size $L=\vert S \vert$ has been divided by $t^{1/3}$. The given parameters: $p/T=0.004$, $n=10$, $\Delta S \in [-3,3]$. Initially, all spins take  the values $\pm 1$ randomly.}
\label{scaling}
\end{figure}

\begin{figure}
\includegraphics[scale=0.3]{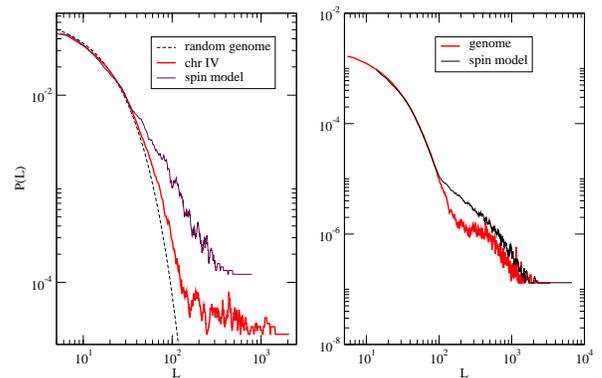}
\caption{Left: distribution of ORF size in chr. IV of the \textit{Saccharomyses cerevisiae} genome and random chromosome, distribution of spin size at $t=5.5 \times 10^6$ MC in a system with parameters as in Fig.~\ref{scaling}.
Right: Same as in the left but for total genome (16 chromosomes) and a system of $N=10^8$ spins, where for $30 \%$ of them  $\Delta S \in [-500,500]$, $p/T=0.004$ and for $70 \%$ of them $\Delta S \in [-3,3]$, $p/T=0.0002$. The presented data have been averaged  in classes of the length of $10$ triplets.}
\label{terrace}
\end{figure}

\begin{figure}
\includegraphics[scale=0.3]{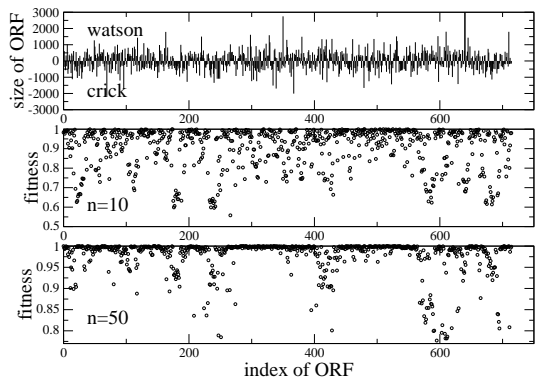}
\caption{Chr. IV of \textit{Saccharomyces cerevisiae} genome: spin representation of ORFs (longer than $150$ triplets), which have been ordered with respect  to location on DNA sequence of their codon START. The corresponding fitness in adapted neighborhood has been plotted for a given $n=10$ nad $n=50$. We have got $<n_0 >  \approx 2.6$ and $<n_0 >  \approx 4.9$, respectively.}
\label{compensation_yeast}
\end{figure}
\begin{figure}
\includegraphics[scale=0.3]{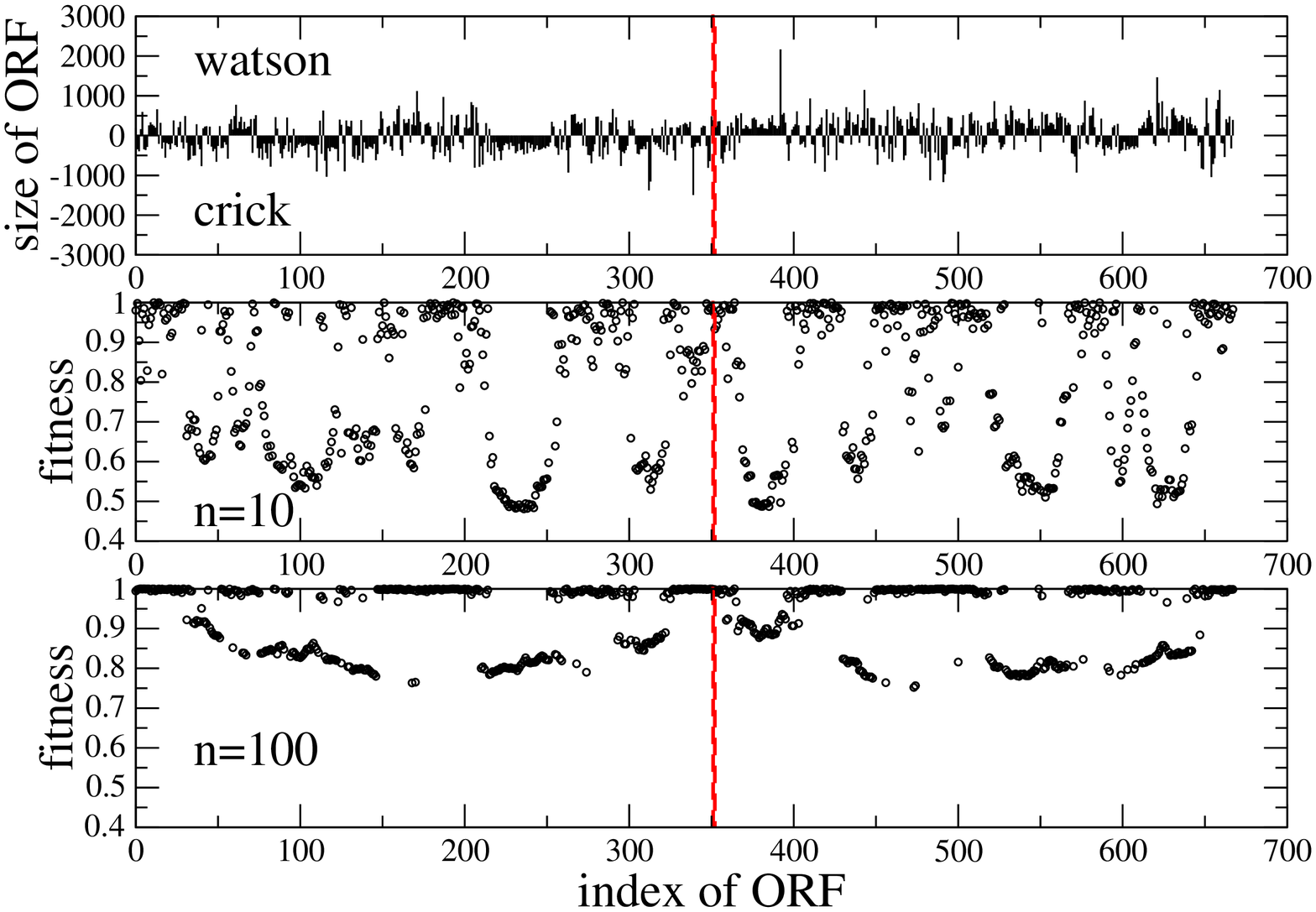}
\caption{\textit{Borrellia burgdorferi} genome: the same as in Fig.~\ref{compensation_borr} but  $n=10$ nad $n=100$. The values of $<n_0 >  \approx 1.3$ and $<n_0 >  \approx 13.2$, respectively. The vertical line has an index of ORF, which is closest to ORI.}
\label{compensation_borr}
\end{figure}

In the left-hand part of Fig.~\ref{scaling}, we have plotted the frequency $P(L)$ of absolute spin values, $L=\vert S \vert$. The right-hand part of the figure is the same but the value $L$ has been rescaled with $t^{1/3}$, where $t$ represents discrete time moment at which the particular distribution $P(L)$ was prepared. One can notice from the figure that these distributions have two distinct wings which correspond to small values of $L$ and large values of $L$. The right wing of the distribution contains the tail-like structure, which develops in time as $t^{1/3}$. This specific shape of the distribution $P(L)$ originates from the long living spin configurations which spontaneously appear in the system after long simulation run. They contain spins 
with the large value of $L$, which are accompanied by spins of opposite sign.  They  are also large, of the order of magnitude of $L$, or there are a few smaller spins compensating the large one. The left wing which corresponds to small 
spins  has been created both by small spins compensating the large ones as well as a large number of small spins representing noise.

It is a situation which has some features common with non-equilibrium experiments on surface diffusion (e.g. \cite{Naumovets}), where adsorbate evolves from an initial overlayer. In these experiments, the time dependence of the diffusing adsorbate is visualized with the help of the concentration profile. The spreading overlayer may undergo series of phase transitions which manifest themselves in a terrace-like concentration profiles. The adsorbate structures represented by these terraces correspond to  domains of new phases which grow with time 
according to the power law 

\begin{equation}
L(t) \sim t^{\alpha}, 
\label{growth_index}
\end{equation}

\noindent
where $L(t)$ represents the linear size of domain and 
$\alpha$ is the growth exponent (see e.g. discussion in \cite{Tringides}).
 This type of physics is a fragment of general domain growth problem \cite{Rutenberg}, where the disordered phase of a physical system is quenched to an ordered phase ($T<T_c$) and the system evolves in time by forming domains of different ordered phases in order to reduce the total surface tension. The large domains grow at the expense of the small ones. Typically, such a system develops a single length scale, $L(t)$, that grows with time as in Eq.(\ref{growth_index}). The growth exponent $\alpha$ is equal to $1/3$  
 for systems with   conserved order parameter and $1/2$ for systems with non-conserved order parameter.
 
In our case, the one-dimensional system of $N$ spins is  very far from   equilibrium, where spins are correlated  according to the self-organization mechanism of the Bak-Sneppen model.  The observed $t^{1/3}$ power law behavior for large spins suggests that the system has properties of a physical system with conserved order parameter.  This is consistent with our choice of the fitness parameter in Eq.(\ref{fitness}). Just the selection for the property  that the value of the sum of all  spins pointing up approaches the absolute value of such sum for the spins pointing down, guarantees this demand.

As pointed out in the Introduction, the distribution of ORF's length in  Fig.~\ref{fig_ORFs}  has two characteristic wings, where the left one represents rather random short sequences, whereas the right one represents long sequences, mostly genes. In Fig.~\ref{terrace}, 
we have compared the distribution $P(L)$ of spin size $L=\vert S \vert$ resulting from our model with the analogous distribution in the natural genome.
Moreover, the results of the analysis of random chromosome of the same nucleotide composition as the chromosome IV of the \textit{Saccharomyces cerevisiae} genome  have been plotted.  The analytical formula for $P(L)$ of
random chromosome has been published in Eq.(7) in  \cite{Dembska}. The reason for the terrace-like shape of the tail part of $P(L)$ in the total genome and the lack of it in the tail of single chromosome which also contributes to this shape, is that the results of 16 chromosomes are presented together.
In order to obtain this terrace-like shape we have assumed 
that there are two groups of spins evolving with different speed. The 
results of the computer simulations have been presented in the right-hand part of the figure.

It is important to remember, that the scaling law, $t^{1/3}$, which we have got for spin evolution, holds only if $\Delta S$ in Eq.(\ref{deltaS}) takes small values and if we can ignore finite size effects. It is always possible to accelerate the computer simulations, even a few times, if  $\Delta S$ is increased. Then, we would have still the $t^{1/3}$ power law with the effect as if we had used a larger time unit. However, in the case of too large a value of $\Delta S$, very quickly the number of small spins becomes too small to  
cover the compensation demand for  $\Delta S$ and the growth process stops rapidly. This is as in grain growth physics, that large grain grows at the expense of smaller grains.

In Figs.~\ref{compensation_yeast} and  \ref{compensation_borr}  we have presented the distribution of genes in the chromosome IV of the \textit{Saccharomyces cerevisiae} genome
and  in the \textit{Borrellia burgdorferi} genome with the help of spin representation. The results for the fitness calculated in an adapted neighborhood for a given value of $n$ have also been plotted. These two figures could be compared with the snapshot of spin configuration and their fitness distribution in our simulations presented in Fig.~\ref{snapshot_spin}. 
We can notice that the results of simulations are closer to those from Fig.~\ref{compensation_yeast}. Only some fragments look very similar in \textit{Borrellia burgdorferi} genome. This observation could also be confirmed by the optimum value of the average  $<n_0>$ for a given $n$ in these examples.
In the case of $n=10$, we have got $<n_0> \approx 2.6$ for \textit{Saccharomyces cerevisiae} genome, $<n_0> \approx 1.3$ for  \textit{Borrellia burgdorferi} genome and 
$<n_0> \approx 2.4$ for the snapshot in Fig.~\ref{snapshot_spin} from the simulations of spin system. 

\section{Conclusions}
We have shown that the specific two-wing distribution of ORF size in the natural genome could originate from the tendency in genome to the balance of the coding sequences in two DNA strands. The consequence of this tendency is that the average gene's length might follow the $t^{1/3}$ power law of evolution.  We have shown this possibility with the help of  the one-dimensional kinetic spin model, in which spins represent length of ORF, and they are correlated in accordance with the Bak-Sneppen model of evolution.

\vspace*{0.2cm}
{\bf Acknowledgments}\\
One of us (MD) thanks Dr. Maria Kowalczuk for discussion on ARS problem in \textit{Saccharomyces cerevisiae} genome. SC was supported by Polish Foundation for Science.

\end{document}